\documentclass[journal]{IEEEtran}
\usepackage{makecell}
\usepackage{setspace}
\usepackage[numbers,sort&compress]{natbib}
\usepackage{amsmath,amssymb,amsfonts}
\usepackage{amsmath}
\usepackage{amsthm}
\usepackage{subcaption}
\usepackage{wasysym}
\usepackage{graphicx}
\usepackage{pifont}  
\usepackage[table]{xcolor}
\usepackage{colortbl}
\usepackage{diagbox}
\usepackage{arydshln}
\usepackage{hyperref}
\usepackage{marvosym}
\usepackage{booktabs,tabulary,multirow,overpic}
\usepackage{caption}
\usepackage[ruled,vlined,linesnumbered]{algorithm2e}
\usepackage[absolute,overlay]{textpos}

\hypersetup{
    colorlinks=true,
    linkcolor=blue,  
    urlcolor=violet,  
    citecolor=blue,
}
\newcommand{\subfigref}[2]{\hyperref[#1]{\ref*{#1}.#2}}
\newtheorem{Definition}{Definition}
\newtheorem{theorem}{Theorem}

\newtheorem{Lemma}{Lemma}
\newtheorem{Corollary}{Corollary}
\newtheorem{Proof}{Proof}
\newtheorem{Proposition}{Proposition}

\newcommand{\Yes}{\CIRCLE}
\newcommand{\No}{\Circle}
\newcommand{\Unknown}{\LEFTcircle}

\makeatletter
\renewcommand\subsection{\@startsection{subsection}{2}{\z@}%
    {1.2ex \@plus 0.1ex \@minus 0.05ex}
    {0.15ex \@plus 0.05ex \@minus 0.03ex}
    {\normalfont\normalsize\itshape}} 
\makeatother

\begin{document}

\title{RoTO: Robust Topology Obfuscation Against Tomography Inference Attacks}

\author{
    \IEEEauthorblockN{Chengze Du\textsuperscript{*}$^1$\thanks{\vspace{-0.5cm}* Equal contribution.}, Heng Xu\textsuperscript{*}$^1$, Zhiwei Yu$^1$, Ying Zhou$^2$, Zili Meng$^3$, Jialong Li$^1$\textsuperscript{\Letter}} \\
    \IEEEauthorblockA{$^1$ Computer Science and Control Engineering, Shenzhen University of Advanced Technology, Shenzhen, China \\
    \textsuperscript{2}School of Electronic Information Engineering, Beijing Jiaotong University\\
    \textsuperscript{3}Department of Electronic and Computer Engineering, Hong Kong University of Science and Technology\\
    }
    \IEEEauthorblockA{\Letter: \texttt{lijialong@suat-sz.edu.cn}}
}

\maketitle
\begin{abstract}
Tomography inference attacks aim to reconstruct network topology by analyzing end-to-end probe delays. Existing defenses mitigate these attacks by manipulating probe delays to mislead inference, but rely on two strong assumptions: (i) probe packets can be perfectly detected and altered, and (ii) attackers use known, fixed inference algorithms. These assumptions often break in practice, leading to degraded defense performance under detection errors or adaptive adversaries. We present RoTO, a robust topology obfuscation scheme that eliminates both assumptions by modeling uncertainty in attacker-observed delays through a distributional formulation. RoTO casts the defense objective as a min-max optimization problem that maximizes expected topological distortion across this uncertainty set, without relying on perfect probe control or specific attacker models. To approximate attacker behavior, RoTO leverages graph neural networks for inference simulation and adversarial training. We also derive an upper bound on attacker success probability, and demonstrate that our approach enhances topology obfuscation performance through the optimization of this upper bound. Experimental results show that RoTO outperforms existing defense methods, achieving average improvements of 34\% in structural similarity and 42.6\% in link distance while maintaining strong robustness and concealment capabilities.
\end{abstract}


\section{Introduction}
\IEEEPARstart{A}{s} an abstract representation of network structure, network topology carries rich information about the physical or logical connections between nodes. It serves as a foundation for network management and security defense~\cite{topo_imp_1, topo_imp_2}. For example, network operation systems rely on network topology information to monitor the status of each link in real time, identify congested paths, and schedule traffic, which helps maintain system stability~\cite{topo_aware_monitor, topo_traffic, congestion_topo, nbt-Duffield_2003, NBT_Duffield_2006, ant-2-Chen_Cao_Bu_2007}. At the same time, security mechanisms such as DDoS traceback and firewall policy deployment also depend heavily on the observability and accuracy of internal topology~\cite{firewall_topo_1, firewall_topo_2, ddos_topo_1, ddos_topo_2}.

From the attacker's perspective, network topology is also a valuable asset. Once the topology is inferred, an attacker can use this information to design precise attack. For instance, by identifying relay nodes or bottleneck links, an attacker can launch destructive link-flooding attacks (LFAs)~\cite{crossfire_attack, g_lfas, linkscope_lfa}; or, by leveraging the topology, they can bypass monitoring nodes to carry out covert communication or data exfiltration~\cite{coremelt_attack}. In systems such as anonymous communication, multi-tenant networks, and censorship-resistant platforms, topology leakage poses particularly severe privacy and security risks~\cite{topo_risk}.

Existing topology inference methods can be roughly divided into two categories: passive and active inference. Passive methods~\cite{Anass_statistical_inference, ye_infer_statstical, bgp, EM2001-passive} include techniques such as traffic correlation analysis, traffic pattern recognition, and routing protocol monitoring. These methods reconstruct the topology by analyzing existing communication data or protocol information in the network. Although they can perform well in certain scenarios, they typically rely on access to traffic logs, exposure of protocol details, or long-term observation of the network. As a result, they are vulnerable to traffic encryption, protocol obfuscation, or policy-based disguise, which limits their effectiveness in modern, security-sensitive networks. 

\begin{figure}[t]
	\centering
	\includegraphics[width=\linewidth]{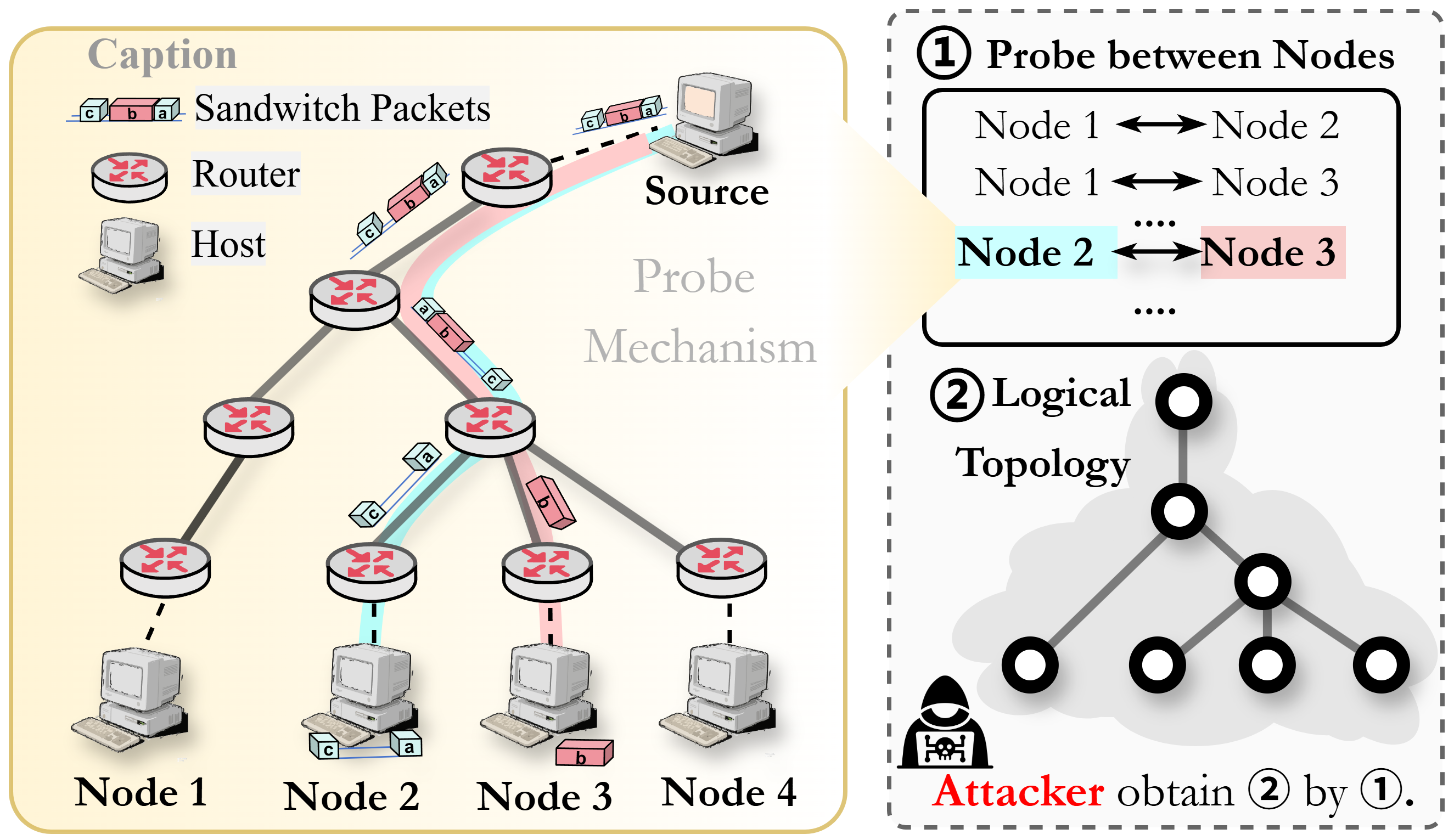}
	\caption{\textbf{Left:} Schematic of the "Sandwitch" packet-sending mechanism for determining shared path information between two leaf nodes. \textbf{Right:} The attacker obtains the final logical topology through extensive probing and subsequent analysis.}
	\label{fig: attack}
 \vspace{-0.5cm}
\end{figure}

In contrast, active methods based on probe packet delays—also known as \textit{Network tomography}—offer a more stealthy approach (a simple schematic diagram is shown in Figure.\ref{fig: attack}). These methods~\cite{crosspoint, mle2002, ni2009efficient, neuralNT, NT-m} send carefully designed probe packets and measure their transmission delays across different network paths. By analyzing the correlation of delays on shared links, they infer the shared bottlenecks and hierarchical structure of the network. Since this approach only requires basic connectivity and does not rely on internal traffic data or protocol information, it can be applied in a wide range of network environments and poses a significant threat.

To defend against topology inference attacks based on active delay measurements, several studies~\cite{antitomo, Proto, secureNT, ChameleonNet} have proposed using delay manipulation techniques to confuse the attacker's observations. These methods aim to interfere with the correlation patterns used in inference by injecting artificial delays or modifying network structure. 

However, these approaches generally rest on \textbf{Two} key assumptions, both of which are fragile in practice and limit the real-world effectiveness of such defenses.

\textit{\textbf{Assumption 1:} \underline{Full control over probe packet delays.}}
Most existing methods assume that the defender can accurately identify probe packets and precisely control their transmission delays. In practice, however, probe packets often experience network congestion, queueing variations, packet loss, or link anomalies, which cause deviations between the actual delays observed by the attacker and the intended delays introduced by the defender~\cite{probe_error_1, probe_error_2, probe_error_3, probe_error_4}. Moreover, intrusion detection systems (IDS) are imperfect in complex traffic environments. False positives and false negatives are common, and difficult to eliminate~\cite{ids-1, ids-2}. Once the attacker receives delay data that do not follow the designed confusion pattern, the entire defense can break down.

\textit{\textbf{Assumption 2:} \underline{Fixed and known attacker models.}}
Many defense mechanisms~\cite{Proto, antitomo, ChameleonNet} are designed with specific inference models in mind, such as shortest-path construction or spectral clustering. However, attackers are not bound to these models. They can adaptively switch algorithms based on observed responses, or apply more robust inference techniques, including graph neural networks~\cite{gnn-nt}. When the actual inference behavior deviates from the defender’s expectations, the confusion effect deteriorates significantly.

To address the strong assumptions in existing methods regarding control over probe packets and attacker modeling, we propose a new topology obfuscation method \textbf{RoTO} (\textbf{Ro}bust \textbf{T}opology \textbf{O}bfuscation). Instead of relying on precise identification of probe packets or fixed attacker models, RoTO adopts an information-theoretic approach. By applying Fano' s inequality~\cite{Fano_inequality}, we derive an upper bound on the attack success probability, which depends on the mutual information between the defender’s perturbation and what the attacker can infer. However, directly minimizing mutual information is challenging. To make the objective tractable, we construct an attacker modeling framework based on a Gibbs posterior~\cite{gibbs-poster}, and convert the problem into minimizing a more practical alternative—expected structural distortion, which measures the expected difference between the topology inferred by the attacker and the ground-truth structure.

To improve robustness, we formulate the defense as an adversarial optimization problem, where the defender designs a perturbation vector to maximize the expected structural error of the attacker under different inference strategies, while keeping the perturbation cost under control. Specifically, we design a min-max optimization framework and integrate graph neural networks (GNNs)~\cite{gnn-review} to learn structure-sensitive perturbation strategies from the real topology. In addition, we introduce a structure-aware stochastic sampling model to better simulate uncertainty in the observation process, ensuring reliable confusion effects under a range of attack conditions.

\textbf{Contribution:} We are the first to propose a unified information theoretic framework for defending against tomography-based inference attacks, which systematically models both the attacker's adaptive inference behavior and the defender's structural obfuscation strategy. Our main contributions are as follows:
\begin{itemize}
    \item \textbf{Theoretical Modeling:} We introduce an information-theoretic formulation and derive an upper bound on the attack success probability using Fano’s inequality. We further transform the defense objective into minimizing the expected structural distortion, which serves as an optimizable surrogate for mutual information.
    \item \textbf{Algorithm Design:} We develop a GNN-based robust perturbation generation module that captures topology-aware structural patterns and design a trainable min-max adversarial optimization framework with structure-aware sampling to enhance the robustness of RoTO against adaptive inference strategies.
    \item \textbf{Experimental Evaluation:} We conduct comprehensive experiments against four representative topology inference attacks on diverse network topologies. Results show that RoTO consistently outperforms existing baseline methods across multiple metrics, achieving up to 34\% improvement, while maintaining superior robustness under noise uncertainty and effective concealment capabilities.
\end{itemize}

The remainder of this paper unfolds our approach systematically. We begin by surveying existing topology obfuscation techniques and their limitations in Section~\ref{sec: related}. Section~\ref{sec: attack} then establishes a realistic threat model that accounts for adaptive adversaries. Building on information-theoretic foundations, Section~\ref{sec: analysis} derives theoretical bounds on attack success probability and formulates our defense objective. These insights guide the design of RoTO in Section~\ref{sec: design}, our robust topology obfuscation framework that integrates graph neural networks with adversarial training. Section~\ref{sec: experiment} presents comprehensive experiments that demonstrate RoTO's effectiveness against diverse inference attacks, followed by concluding remarks and future directions in Section~\ref{sec: conclusion}.

\vspace{-0.1in}
\section{Related Work}
\label{sec: related}

\textbf{Topology inference} aims to uncover hidden network structures using traffic features or behavior.Traffic correlation-based methods rely on the idea that paths sharing physical resources show similar characteristics. Some approaches~\cite{Anass_statistical_inference, ye_infer_statstical} used statistical dependency analysis, and CrossPoint~\cite{crosspoint} combined delay covariance with spectral clustering~\cite{spectral_cluster}. NetFlow~\cite{NetFlow} based methods~\cite{NetFlow_Analyzer} extract time patterns, while mutual information~\cite{MIR_timeseries} captures nonlinear flow dependencies, effective even in anonymous networks. Statistical delay analysis detects shared congestion via variance~\cite{delay_shared_congestion_localization, topo_infer_by_delay}, and clock skew reveals topology through timing differences~\cite{clock_syn}. Application-layer traffic analysis uses deep packet inspection and machine learning classifiers to detect topology-related patterns~\cite{DIP_ML, DIP_DL}. Flow correlation attacks~\cite{flow_cor_attack} exploit timing in encrypted traffic to rebuild communication graphs, even in Tor. Protocol-based techniquesinclude BGP monitoring~\cite{bgp, bgp-2} and OSPF analysis~\cite{ospf} for mapping inter- and intra-domain structures. Deep learning further improves inference by learning node-level features~\cite{gnn-nt, neuralNT}.

\textbf{Defenses} against topology inference operate across layers and target different attack strategies. Traffic obfuscation is the most widely used category. It includes dummy traffic injection (such as padding cells in Tor)~\cite{10.1145/2534169.2486002, flow_inject, traffic_inject}, traffic shaping through rate limiting and packet normalization~\cite{traffic_shape, traffic_shape_2}, transformation-based obfuscation that mimics other applications, and adaptive padding that adjusts parameters based on attack patterns~\cite{Protocol_Obfuscation_GRINStudy, auto_muffler}. Path diversity techniques~\cite{MPTCP, defenseshortestpathattacks, HOHO, UCMP} aim to break stable-path assumptions using onion routing with layered encryption and random selection, multipath protocols that split flows across routes, and path length randomization by adding dummy hops. Resource allocation defenses~\cite{DBA4DOS} adjust link capacity dynamically, though practical deployment remains limited due to performance trade-offs. Protocol-level defenses~\cite{NetHide, SpiderNet, decorrelation} include encrypted tunnels, protocol mimicry that disguises traffic as common protocols, and timing obfuscation with calibrated delays to reduce correlation.

\begin{figure}[t]
	\centering
	\includegraphics[width=\linewidth]{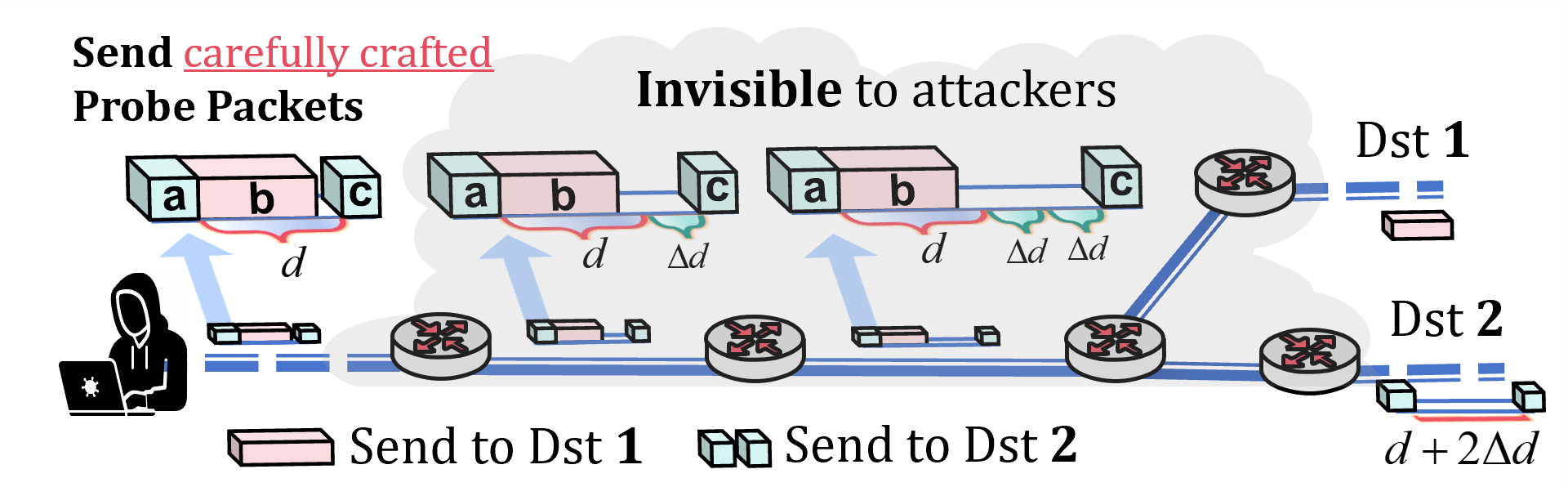}
	\caption{Attackers infer the number of shared links (or other key messages) between Destination 1, Destination 2, and the source by observing the increased delay between packets \texttt{a} and packet \texttt{c} (caused by packet \texttt{b}'s interference).}
	\label{fig: attack_detail}
 \vspace{-0.4cm}
\end{figure}

Network tomography~\cite{nbt-Duffield_2003, neuralNT} infers internal topology from end-to-end delay measurements, without needing access to intermediate nodes or routing data (see in Figure~\ref{fig: attack_detail}). Early methods like MLE~\cite{mle2002} and EM~\cite{EM2001-passive} use additive delay models to solve inverse problems and locate shared bottlenecks. Algebraic techniques such as SLE~\cite{ni2009efficient} handle sparse measurements. Recent deep learning approaches~\cite{gnn-nt} learn topology-delay mappings directly from observations. These attacks require only basic connectivity, without internal access or protocol knowledge. Defenses focus on delay manipulation: AntiTomo~\cite{antitomo}, ChameleonNet~\cite{ChameleonNet} and Proto~\cite{Proto} inject artificial delays; EigenObfu~\cite{Eigenobfu} modifies graph structures; SecureNT~\cite{secureNT} takes into account the impact on benign probing. However, most rely on detectability or static threat models, reducing their robustness against adaptive attacks.

\section{Threat Model}
\label{sec: attack}
\begin{table}[t]
\centering
\caption{\footnotesize \MakeUppercase{Comparison of Topology Obfuscation Methods Against Network Tomography Attacks.}}
\renewcommand{\arraystretch}{0.7}
\small
\begin{tabular}{lcccc}
\hline\toprule
\textbf{Method} & \makecell[c]{Deploy-\\ability} & \makecell[c]{Stealth-\\iness} & \makecell[c]{Key Info\\ Hidden} & \makecell[c]{Threat\\ Model} \\
\midrule
EigenObfu~\cite{Eigenobfu}     & \No      & \Unknown & \Yes      & -                           \\
Proto~\cite{Proto}         & \Yes     & \No      & \No       & \cite{mle2002} \\
AntiTomo~\cite{antitomo}      & \Yes     & \Yes     & \No       &\cite{ni2009efficient} \\
SecureNT~\cite{secureNT}     & \Yes      & \No     & \No      &\cite{mle2002}                           \\
\makecell[l]{\footnotesize ChameleonNet~\cite{ChameleonNet}}  & \Yes     & \Yes     & \Yes       & \cite{ni2009efficient} \\
\midrule
\textbf{Proposal} (RoTO) & \Yes     & \Yes     & \Yes      & \textbf{Diversity} \\
\bottomrule[1pt]
\end{tabular}
\label{tab:obfuscation_comparison}
\vspace{-0.4cm}
\end{table}

This section establishes the threat model for RoTO, defining the network environment, adversary capabilities, and defender constraints. Unlike existing work that assumes perfect probe packet detection and fixed attacker models, we adopt a more realistic modeling approach that accounts for practical limitations in real-world deployments.

\subsection{Network Model}
We consider a tree-structured network $\mathcal{T} = (V, E)$ rooted at a source node $s$, with $l$ leaf (receiver) nodes. The network topology is encoded by a binary matrix $A \in \mathbb{A} \subset \{0,1\}^{k \times m}$, where $k = \binom{l}{2}$ represents all receiver node pairs and $m$ is the number of internal links. Let $\mu \in \mathbb{R}^{m}$ denote the link delay vector. The end-to-end delay vector observed by the attacker is:
\begin{equation}
    X = A\mu.
\end{equation}

To obfuscate the inference process, the defender perturbs $X$ to generate $\tilde{X}$, which is further affected by adversarial uncertainty to produce the final observation $X^* \sim \mathcal{D}(\tilde{X})$. The attacker then infers the topology $\hat{A} = f_{\text{adv}}(X^*)$. Our goal is to design $\tilde{X}$ such that $\hat{A} \neq A$ with high probability and high structural distortion.

\begin{theorem}[Topology-Delay Injectivity]. 
\label{thm: inj}
Given a known link delay vector $\mu$, the mapping $f: A \mapsto A\mu$ from tree topology $A$ to shared path matrix $X$ is injective. That is, for any two distinct topologies $A_1 \neq A_2$, we have $A_1\mu \neq A_2\mu$.
\end{theorem}

\textbf{Proof} of Theorem~\textbf{\ref{thm: inj}} is provided in Appendix~\ref{sec: appendix}. $\hfill \blacksquare$

\begin{Corollary} 
This follows directly from Theorem~\ref{thm: inj}. If an attacker obtains precise observations $X$ and knows the link delay vector $\mu$, the true topology $A$ can be uniquely recovered \emph{in principle}. This implies that each component of $X$ carries structural information about the network topology.
\end{Corollary}

This injectivity property reveals the fundamental sensitivity of network tomography: any perturbation to the shared path matrix may affect topology reconstruction results, providing theoretical justification for our obfuscation approach. Note that while mathematical injectivity guarantees the existence of a unique solution, practical inference algorithms may only approximate the optimal solution due to computational limitations and heuristic approaches, creating a gap between theoretical recoverability and practical inference capability.

\subsection{Adversary Capability}
Unlike previous work that assumes deterministic inference (e.g., RNJ~\cite{ni2009efficient}, MLE~\cite{mle2002}), we consider a general class of active inference attackers who aim to reconstruct the underlying routing topology $A$ based on observed end-to-end delays $X^*$. Unlike prior works that assume a fixed inference function, we model the attacker as optimizing over a strategy space $\Phi$, which includes both classical tomography algorithms and adaptive, data-driven approaches.

Given the noisy delay observation $X^*$, the attacker selects an inference function $f_{\text{adv}} \in \Phi$ to produce an estimate $\hat{A} = f_{\text{adv}}(X^*)$. The attacker seeks to minimize reconstruction loss relative to the true topology:
\begin{equation}
    f^*_{\text{adv}} = \arg \min_{f \in \Phi} \ \mathbb{E}[d(A, f(X^*))]
\end{equation}

Here, $d(\cdot, \cdot)$ is a structural distance metric over topology space (e.g., Hamming distance over adjacency matrices). This formulation allows the attacker to adaptively update inference strategies in response to environmental conditions or observed perturbation patterns. The defender must therefore anticipate the worst-case behavior over the entire function class $\Phi$.

\subsection{Defender Model}
The defender operates under the constraint that probe packet detection and delay manipulation are imperfect in practice. We model this imperfection through the uncertainty distribution $\mathcal{D}(\tilde{X})$, which captures the gap between the defender's intended perturbation $\tilde{X}$ and what the attacker actually observes $X^*$. This uncertainty arises from several sources: network congestion causing unpredictable delays, imperfect intrusion detection systems producing false positives and negatives, and packet loss or routing anomalies. The defender's objective is to design perturbations that remain effective even under such uncertainty. Rather than assuming perfect control over probe packets, the defender must account for the distributional nature of the final observations and optimize for robustness across the entire uncertainty set. This approach acknowledges the practical limitations of real-world deployments while maintaining theoretical rigor in the defense design.

\section{Theoretical Foundation}
\label{sec: analysis}
This section develops the theoretical foundation of RoTO from an information-theoretic perspective. We derive an upper bound on attack success probability using Fano's inequality~\cite{Fano_inequality} and formulate a tractable defense objective based on expected structural divergence.
\begin{table}[t]
\raggedright %
\caption{\footnotesize{LIST OF KEY SYMBOLS.}}
\label{tab:notation}
\small
\begin{tabular}{ll}
\hline\toprule
\textbf{Description} & \textbf{Symbol} \\
\midrule
Structured network with vertices and edges & $T = (V, E)$ \\
Binary matrix encoding topology & $A \in \{0,1\}^{k \times m}$ \\
Link delay vector & $\mu \in \mathbb{R}^m$ \\
End-to-end delay observed by attacker & $X = A\mu$ \\
Perturbed delays manipulated by defender& $\tilde{X}$ \\
Attacker's observation affected by uncertainty & $X^* \sim \mathcal{D}(\tilde{X})$ \\
Inferred topology by attacker & $\hat{A} = f_{adv}(X^*)$ \\
Uncertainty distribution modeling noise & $\mathcal{D}(\cdot)$ \\
Gibbs posterior distribution & $P_\beta(A'|X^*)$ \\
Structural distance metric & $d(A, A')$ \\
Attack success probability & $P_{succ}$ \\
Expected structural divergence & $\mathcal{D}_{struct}(\tilde{X})$ \\
GNN perturbation function & $\pi_\theta(G)$ \\
MI between true topology and observation & $I(A; X^*)$ \\
\bottomrule[1pt]
\end{tabular}
\vspace{-0.4cm}
\end{table}

\subsection{Attack Success Probability Upper Bound}
\label{sub: theory-upper}
We aim to quantify the probability that an adversary successfully reconstructs the true network topology. Let $A \sim \mathcal{U}(\mathbb{A})$ denote the true topology uniformly sampled from the feasible set $\mathbb{A}$. Let $X^*$ represent the attacker's observed data, and $\hat{A} = f_{\text{adv}}(X^*)$ be the inferred topology via adversarial inference function $f_{\text{adv}}$.

\begin{Definition}[Attack Success Probability]
The \emph{attack success probability} is defined as the probability that the adversary exactly recovers the true topology:
\begin{equation}
    P_{\mathrm{succ}} := \Pr[\hat{A} = A]
\end{equation}
\end{Definition}

We now present an upper bound on this quantity based on information-theoretic principles.

\begin{theorem}[Fano-based Upper Bound]
\label{thm: upper-bound}
Let $I(A; X^*)$ denote the mutual information between the true topology $A$ and the attacker's observation $X^*$. Then the attack success probability is bounded by:
\begin{equation}
    P_{\mathrm{succ}} \leq \frac{I(A; X^*) + 1}{\log |\mathbb{A}|}
\end{equation}
\end{theorem}

\begin{Proof}
Let $P_e := \Pr[\hat{A} \neq A]$ denote the error probability of any estimator of $A$ based on $X^*$. According to Fano's inequality, we have:
\begin{equation}
    H(A \mid X^*) \leq h(P_e) + P_e \log(|\mathbb{A}| - 1)
\end{equation}
where $h(p) = -p \log p - (1 - p) \log(1 - p)$ is the binary entropy function. Since $h(P_e) \leq 1$ for all $P_e \in [0, 1]$, we obtain:
\begin{equation}
    H(A \mid X^*) \leq 1 + P_e \log |\mathbb{A}|
\end{equation}
Rearranging yields:
\begin{equation}
    P_e \geq \frac{H(A \mid X^*) - 1}{\log |\mathbb{A}|}
\end{equation}
Noting $P_{\mathrm{succ}} = 1 - P_e$, we obtain:
\begin{equation}
    P_{\mathrm{succ}} \leq 1 - \frac{H(A \mid X^*) - 1}{\log |\mathbb{A}|}
\end{equation}
Using $I(A; X^*) = H(A) - H(A \mid X^*)$ and $H(A) = \log |\mathbb{A}|$, we substitute to get:
\begin{equation}
    P_{\mathrm{succ}} \leq \frac{I(A; X^*) + 1}{\log |\mathbb{A}|} \tag*{$\blacksquare$}
\end{equation}
\end{Proof}

\subsection{Expected Structural Divergence}

The mutual information $I(A; X^*)$ in \textbf{Theorem}~\textbf{\ref{thm: upper-bound}} provides a principled upper bound on attack success probability, but it is computationally intractable to optimize directly. To address this challenge, we introduce a surrogate objective based on expected structural divergence that serves as a practical approximation while maintaining theoretical rigor.

\begin{Definition}[Expected Structural Divergence]
\label{def: esd}
Given a perturbed observation $\tilde{X}$ and uncertainty distribution $\mathcal{D}(\cdot)$, the expected structural divergence is defined as:
\begin{equation}
    \mathcal{D}_{\text{struct}}(\tilde{X}) := \mathbb{E}_{X^* \sim \mathcal{D}(\tilde{X})} \left[ \mathbb{E}_{A' \sim f_{adv}( X^*)} \left[ d(A, A') \right] \right]
\end{equation}
where $d(A, A')$ is a structural distance metric between topologies (e.g., Hamming distance on edge sets or tree edit distance), and $f_{adv}$ represents the attacker's inference strategy.
\end{Definition}

Intuitively, $\mathcal{D}_{\text{struct}}(\tilde{X})$ quantifies the expected error in the attacker's topology reconstruction under the defender's perturbation strategy. The key insight is to establish a connection between this structural divergence and the conditional entropy $H(A \mid X^*)$ that appears in our attack success probability bound. The following proposition connection allows us to transform the information-theoretic objective into a more tractable optimization problem.

\begin{Proposition}
\label{prop: structural_d}
Under bounded structural diameter $\max_{A, A'} d(A, A') \leq D$ and distance-sensitive Gibbs posterior $P_\beta(A' \mid X^*) \propto e^{-\beta d(A, A')}$, the conditional entropy satisfies:
\begin{equation}
    H(A \mid X^*) \geq \frac{1}{C_d} \cdot \mathbb{E}_{A' \sim P_\beta(A' \mid X^*)} \left[d(A, A')\right] - \log C_d
\end{equation}
where $C_d = \frac{\log |\mathbb{A}|}{\beta}$ is a constant depending on the inverse temperature parameter $\beta$ and the topology space $\mathbb{A}$.
\end{Proposition}

\begin{Proof}
Let $A$ be the true adjacency matrix and $\mathbb{A}$ be the graph topology space with diameter bounded by $D$. Consider the Gibbs posterior distribution:
\begin{equation}
P_\beta(A' \mid X^*) = \frac{\exp(-\beta \cdot d(A, A'))}{Z}
\end{equation}

where the partition function is:
\begin{equation}
Z := \sum_{A'' \in \mathbb{A}} \exp(-\beta \cdot d(A, A''))
\end{equation}

The conditional entropy can be written as:
\begin{equation}
H(A \mid X^*) = -\sum_{A' \in \mathbb{A}} P_\beta(A' \mid X^*) \log P_\beta(A' \mid X^*)
\end{equation}

Substituting the Gibbs posterior form:
\begin{align}
H(A & \mid X^*) 
= -\sum_{A'} \left( \frac{e^{-\beta d(A, A')}}{Z} \right) 
    \cdot \log \left( \frac{e^{-\beta d(A, A')}}{Z} \right) \\
&= -\sum_{A'} \left( \frac{e^{-\beta d(A, A')}}{Z} \right) 
    \cdot \left( -\beta d(A, A') - \log Z \right)  \\
&= \sum_{A'} \left( \frac{e^{-\beta d(A, A')}}{Z} \right) 
    \cdot \left( \beta d(A, A') + \log Z \right)  \\
&= \beta \cdot \sum_{A'} \left( 
        \frac{e^{-\beta d(A, A')}}{Z} \cdot d(A, A') 
    \right) \nonumber \\
&\quad + \log Z \cdot \sum_{A'} \left( 
        \frac{e^{-\beta d(A, A')}}{Z} 
    \right) \\
&= \beta \cdot \mathbb{E}_{A' \sim P_\beta} 
    \left[ d(A, A') \right] + \log Z
\end{align}

Since the structural distance is bounded, $0 \leq d(A, A') \leq D$, we have:
$
e^{-\beta D} \leq e^{-\beta d(A, A')} \leq 1
$.
For the partition function, we can establish an upper bound:
\begin{equation}
Z = \sum_{A'' \in \mathbb{A}} e^{-\beta d(A, A'')} \leq \sum_{A'' \in \mathbb{A}} 1 = |\mathbb{A}|
\end{equation}

Therefore, we obtain:
$
\log Z \leq \log |\mathbb{A}|
$.

From the entropy expansion, we have:
\begin{equation}
H(A \mid X^*) = \beta \cdot \mathbb{E}_{A' \sim P_\beta}[d(A, A')] + \log Z
\end{equation}

Using the upper bound $\log Z \leq \log |\mathbb{A}|$:
\begin{equation}
H(A \mid X^*) \geq \beta \cdot \mathbb{E}[d(A, A')] - \log |\mathbb{A}|
\end{equation}

Setting $C_\beta := \frac{\log |\mathbb{A}|}{\beta}$ (under the assumption $\beta \leq \log |\mathbb{A}|$ to ensure $C_\beta \geq 1$):
\begin{equation}
H(A \mid X^*) \geq \frac{1}{C_\beta} \cdot \mathbb{E}[d(A, A')] - \log C_\beta
\end{equation}

This completes the Proof with $C_d = C_\beta = \frac{\log |\mathbb{A}|}{\beta}$. $\hfill\blacksquare$
\end{Proof}

\textbf{Implications.} Combined with \textbf{Theorem~\ref{thm: upper-bound}}, \textbf{Proposition~\ref{prop: structural_d}} implies that maximizing the expected structural divergence $\mathcal{D}_{\text{struct}}(\tilde{X})$ effectively increases the conditional entropy $H(A \mid X^*)$, which in turn reduces the mutual information $I(A; X^*)$ and consequently decreases the attack success probability upper bound.

\subsection{Final Defense Objective}

Building on the theoretical results above, we now formulate the final defense objective for the defender. However, in practice, the defender must also consider the cost of perturbation and the adaptivity of the attacker.

To this end, we introduce a perturbation cost function:
\begin{equation}
C(\tilde{X}, X) := \|\tilde{X} - X\|_2^2,
\end{equation}
This ensures that the defender introduces only bounded overhead to the system.

Furthermore, we consider a more robust defense model where the attacker's reasoning strategy $f_{\text{adv}}$ is no longer fixed but can adaptively choose behaviors that maximize attack effectiveness. In this case, the defender's optimization objective becomes the following adversarial min-max problem:
\begin{equation}
\min_{\tilde{X}} \max_{f_{\text{adv}} \in \Phi} \mathbb{E}_{X^* \sim \mathcal{D}(\tilde{X})} \left[ \mathbb{E}_{A' \sim f_{\text{adv}}(X^*)} [d(A, A')] \right] + \lambda \cdot C(\tilde{X}, X)
\label{eq: defend-obj}
\end{equation}

The defender's goal is to find the most robust perturbation set $\mathcal{D}(\tilde{X})$ such that any attacker strategy finds it difficult to recover the original structure.

\section{RoTO Design}
\label{sec: design}
\begin{figure*}[t]
	\centering
	\includegraphics[width=\textwidth]{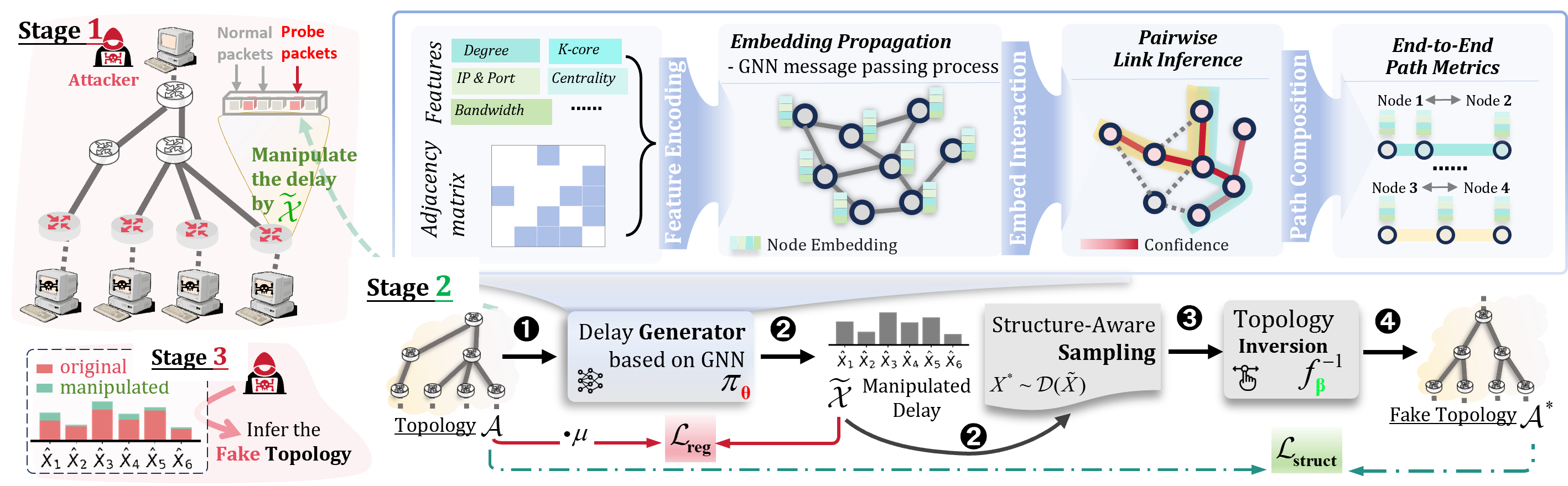}
         \caption{Overview of the \textbf{RoTO} Framework which illustrates the three-stage workflow. \textbf{Stage 1} depicts attackers injecting probe packets to infer network topology via end-to-end delays. \textbf{Stage 2} shows the GNN-based delay generator ($\pi_\theta$) manipulating delays and generating perturbed delays ($\tilde{X}$). \textbf{Stage 3} demonstrates the structure-aware sampling process ($X^* \sim \mathcal{D}(\tilde{X})$) that introduces uncertainty into attacker observations, followed by topology inversion ($f_\beta^{-1}$) to infer a distorted topology. }
	\label{Overview}
         \vspace{-0.5cm}
\end{figure*}

In this section, we present the design of the RoTO defense framework, which integrates topology-aware perturbation generation~\S\ref{subsec: design-gnn}, structure-consistent noise modeling~\S\ref{subsec: design-sampling}, and adversarially training~\S\ref{subsec: adv-train} against probabilistic inference attacks~\S\ref{subsec: design-topoinfer}. The overall framework is shown in Figure~\ref{Overview}.
\subsection{Overview}

The core idea of RoTO is to proactively distort attacker-visible end-to-end delays by injecting graph-structured perturbations, such that the attacker’s inferred topology diverges significantly from the true structure, even in the presence of adaptive inference strategies and observation uncertainty.

The defense pipeline comprises three tightly coupled stages. First, given the ground-truth network topology {\large \ding{182}}, RoTO leverages a graph neural network (GNN) to learn topology-aware delay perturbations. This GNN takes the original tree topology as input and outputs a modified delay {\large \ding{183}} vector that induces structural confusion while controlling perturbation cost. Second, to reflect realistic uncertainty in attacker observation (e.g., due to congestion or IDS inaccuracies), the perturbed delay vector {\large \ding{184}} is further sampled through a structure-aware noise model to produce integer-valued observations that are consistent with tree metrics. Finally, these noisy delay signals are processed by a simulated attacker modeled via a Gibbs posterior distribution, which yields a probabilistic estimate of the underlying fake topology {\large \ding{185}}. The defender is trained to maximize the expected structural error in the inferred topology by optimizing the GNN under adversarial training, thus learning perturbations that are robust across a range of attacker strategies.

\subsection{GNN-based Delay Perturbation Generator}
\label{subsec: design-gnn}
To implement the perturbation strategy $\pi_\theta$, we design a structure-sensitive graph neural network (GNN) that takes the true network topology as input and outputs a perturbed delay vector $\widetilde{X} \in \mathbb{Z}^k$, where each entry $\widetilde{X}_{ij}$ denotes the defender-specified shared delay between leaf node pair $(i, j)$. The GNN operates over the attributed tree graph $G = (V, E)$, where leaf nodes represent receivers, and internal nodes model branching points in the routing hierarchy.

Each node $v \in V$ is initialized with a feature vector $h^{(0)}_v$. For leaf nodes, we use one-hot encoding to capture their identity; for internal nodes, we use learnable embeddings. Each edge $(u, v) \in E$ carries associated features $e_{uv}$, including the link delay $\mu_{uv}$ and the subtree size or depth if available.

We adopt a message-passing GNN framework tailored for trees, where node representations are iteratively updated via:

\begin{equation}
    \begin{aligned}
        &m^{(l+1)}_{u \to v} = \text{MSG}^{(l)}\left(h^{(l)}_u, h^{(l)}_v, e_{uv}\right) \\
        &h^{(l+1)}_v = \text{UPDATE}^{(l)}\left(h^{(l)}_v, \text{AGG}^{(l)}\left(\{ m^{(l+1)}_{u \to v} \mid u \in \mathcal{N}(v) \}\right)\right)
    \end{aligned}
\end{equation}

where $\mathcal{N}(v)$ denotes the child nodes of $v$ in the tree, and $\text{MSG}^{(l)}$, $\text{UPDATE}^{(l)}$, and $\text{AGG}^{(l)}$ are learnable functions implemented as multi-layer perceptrons (MLPs) and pooling operators (e.g., sum or mean).

After $L$ layers of message passing, each node $v$ obtains a final embedding $h^{(L)}_v$. For each receiver pair $(i, j)$, we compute their pairwise representation using a readout function $\phi$, such as concatenation or Hadamard product:
$ z_{ij} = \phi(h^{(L)}_i, h^{(L)}_j)$.
   
The overall perturbation output is a symmetric matrix $\widetilde{X} \in \mathbb{Z}^{l \times l}$, satisfying:
\begin{equation}
    \begin{aligned}
        \widetilde{X} = \pi_\theta(G) \Longleftarrow \widetilde{X}_{ij} = \text{MLP}_\theta(z_{ij})
    \end{aligned}
\end{equation}
where the $\widetilde{X}_{ij}$ is then generated by applying an MLP over the pairwise feature, $l$ denotes the number of leaf nodes. The matrix is vectorized into a $k$-dimensional vector with $k = \binom{l}{2}$, corresponding to all receiver node pairs. This output serves as input to the downstream stochastic sampling module and attacker inference process.

By training the GNN under an adversarial loss (detailed later), we ensure that $\pi_\theta$ learns to maximize structural distortion in the attacker’s reconstruction, while maintaining realistic and resource-constrained delay manipulation.

\subsection{Structure-Aware Sampling}
\label{subsec: design-sampling}
In practice, the perturbed delays designed by the defender are further affected by environmental uncertainty, such as network congestion, queueing delay, and path-level jitter. Importantly, such uncertainty is unidirectional: due to physical and protocol-level constraints, delays can increase or remain constant, but they cannot decrease below the defender’s intended values. To faithfully capture this asymmetry, we model the attacker’s observations as the result of additive positive Gaussian noise applied to the defender's delay output.

Formally, let $\widetilde{X} \in \mathbb{R}^k$ denote the shared path delay vector generated by the GNN. The attacker-observed delay vector $X^*$ is sampled from a structure-consistent positive-noise distribution $D_\epsilon(\cdot)$ as:
\begin{equation}
    X^* \sim \mathcal{D}(\tilde{X})  = \widetilde{X} + \Delta, \quad \Delta \sim \text{ReLU}(\mathcal{N}(0, \epsilon^2 I))
\end{equation}
where $\mathcal{N}(0, \epsilon^2 I)$ denotes an isotropic zero-mean Gaussian distribution and $\text{ReLU}(\cdot)$ is applied element-wise to ensure all noise terms are non-negative. That is, for each element $(i, j)$:
\begin{equation}
    X^*_{ij} = \widetilde{X}_{ij} + \max\{0, \delta_{ij}\}, \quad \delta_{ij} \sim \mathcal{N}(0, \epsilon^2)
\end{equation}

This formulation reflects the realistic bias of delay perturbation toward inflation, consistent with the fact that congestion or queuing delays are inherently additive. Moreover, it maintains continuity and differentiability, which are desirable for gradient-based optimization during training.

To ensure that the attacker’s observed delays remain embeddable in a valid tree metric, we further impose an ultrametric constraint~\cite{gnn-nt} on $X^*$. For any triple of leaf nodes $i, j, k$, the sampled delays must satisfy:
\begin{equation}
    X^*_{ij} \leq \max\{X^*_{ik}, X^*_{jk}\}
\end{equation}

This constraint is enforced through post-sampling projection onto the ultrametric space, ensuring that attacker observations remain topologically consistent and can be mapped to a plausible tree structure. The structure-aware noise model thereby simulates realistic, noisy observations while preserving the feasibility of topology inference.

\subsection{Gibbs Posterior-Based Topology Inference}
\label{subsec: design-topoinfer}
To simulate an adaptive and probabilistic attacker, we adopt a Bayesian formulation in which the adversary maintains a belief distribution over candidate topologies conditioned on noisy observations $X^*$. Rather than committing to a deterministic inference algorithm, the attacker selects from a space of plausible topologies using a Gibbs posterior, which biases toward structures that better explain the observed delays.

Formally, let $\mathcal{A}$ be the set of all feasible tree topologies over $l$ receivers, and let $\mu \in \mathbb{R}^m$ denote the link delay vector. For each candidate topology $A' \in \mathcal{A}$, the attacker evaluates its fitness using the squared loss:
\begin{equation}
    \mathcal{L}(X^*, A') := \|X^* - A' \mu\|^2_2
\end{equation}

The posterior distribution over topologies is then given by:
\begin{equation}
    P_\beta(A' \mid X^*) = \frac{1}{Z(X^*)} \exp\left( -\beta \cdot \mathcal{L}(X^*, A') \right)
\end{equation}
where $\beta > 0$ is the inverse temperature parameter controlling the sharpness of the distribution, and $Z(X^*)$ is the partition function:
\begin{equation}
    Z(X^*) = \sum_{A'' \in \mathcal{A}} \exp\left( -\beta \cdot \mathcal{L}(X^*, A'') \right)
\end{equation}

This formulation reflects a continuum of attacker behaviors. When $\beta \to \infty$, the attacker behaves deterministically, choosing the MAP topology $A^* = \arg\min_{A'} \mathcal{L}(X^*, A')$. When $\beta \to 0$, the attacker becomes maximally uncertain, assigning uniform weight to all candidates.

Under this posterior, the attacker’s expected structural accuracy is quantified by the expected structural similarity between the ground-truth topology $A$ and the inferred topology $A'$, measured via a distance metric $d(A, A')$, such as the tree edit distance or Hamming distance over edge indicators. 
\begin{equation}
    \mathbb{E}_{A' \sim P_\beta(\cdot \mid X^*)}[d(A, A')]
\end{equation}
This quantity serves as the optimization target for the defender, which aims to maximize this expected distance, thereby minimizing the attacker's inference accuracy.

By integrating this probabilistic inference model with our perturbation and sampling modules, we are able to train the defender against a broad class of adaptive inference strategies, rather than overfitting to any specific attack algorithm.

\subsection{Training Objective and Adversarial Optimization}
\label{subsec: adv-train}

To optimize the perturbation generator $\pi_\theta$ under adversarial conditions, we formulate the defender’s training objective as a min-max problem that balances structural misleading effectiveness and perturbation efficiency. The core idea is to find a perturbation strategy that maximizes the expected structural error of the attacker—regardless of the attacker’s specific inference behavior—while constraining the distortion introduced into the network.

As derived in Section~\ref{sub: theory-upper}, the attacker’s inference accuracy is upper-bounded by a function of the mutual information between the true topology $A$ and the attacker’s observation $X^*$. Since direct minimization of mutual information is intractable, we adopt the expected structural divergence $D_{\text{struct}}(\widetilde{X})$ as a surrogate objective:
\begin{equation}
    D_{\text{struct}}(\widetilde{X}) := \mathbb{E}_{X^* \sim D(\widetilde{X})} \left[ \mathbb{E}_{A' \sim P_\beta(\cdot \mid X^*)} \left[ d(A, A') \right] \right]
\end{equation}

To operationalize this in learning, we define the structural misleading loss as the negative expected divergence:
\begin{equation}
    \mathcal{L}_{\text{struct}}(\theta, \beta) := - \mathbb{E}_{X^* \sim D(\pi_\theta(G))} \left[ \mathbb{E}_{A' \sim P_\beta(\cdot \mid X^*)} \left[ d(A, A') \right] \right]
\end{equation}
where $\pi_\theta(G)$ denotes the perturbed delay vector produced by the GNN for topology graph $G$, and $\beta$ captures the attacker's inference sharpness.

\begin{algorithm}[t]
    \small
    \setstretch{1}
    \SetKwInOut{Input}{Input}
    \SetKwInOut{Output}{Output}
    \caption{RoTO's Workflow}
    \label{alg:roto}
    \Input{True topology $G = (V, E)$, link delays $\mu$, \\
    distance metric $d(\cdot, \cdot)$, \\
    attacker confidence range $[\beta_{\min}, \beta_{\max}]$}
    \Output{Trained perturbation generator $\pi_\theta$}
    \BlankLine
    \tcp{\color{blue}\small{\textbf{Defender}}}
    Compute ground-truth delay vector: $X \leftarrow A\mu$\;
    Generate perturbed delays: $\widetilde{X} \leftarrow \pi_\theta(G)$\;
    \BlankLine
    Add positive noise: $X^* \leftarrow \widetilde{X} + \max(0, \mathcal{N}(0, \epsilon^2))$\;
    Project $X^*$ onto ultrametric space to ensure tree consistency\;
    \BlankLine
    \tcp{\color{blue}\small{\textbf{Sim-Attacker}}}
    \For{$\beta \in [\beta_{\min}, \beta_{\max}]$}{
        Compute topology posterior:\;
        \Indp $P_\beta(A' \mid X^*) = \frac{1}{Z(X^*)} \exp\left( -\beta \cdot \|X^* - A'\mu\|^2 \right)$\;
        Evaluate expected structure loss:\;
        \Indp $\mathcal{L}_{\text{struct}}(\theta, \beta) = -\mathbb{E}_{A' \sim P_\beta}[d(A, A')]$\;
        \Indm
    }
    Select worst-case attacker parameter: $\beta^* \leftarrow \arg\max_\beta \mathcal{L}_{\text{struct}}(\theta, \beta)$\;
    \BlankLine
    \tcp{\color{blue}\small{\textbf{Adversarial Optimization}}}
    Compute regularization: $\mathcal{L}_{\text{reg}}(\theta) = \|\widetilde{X} - X\|_2^2$\;
    Update $\theta$ to minimize: $\mathcal{L}_{\text{struct}}(\theta, \beta^*) + \lambda \cdot \mathcal{L}_{\text{reg}}(\theta)$\;
    \BlankLine
    \Return{Trained $\pi_\theta$}
\end{algorithm}

To prevent the defender from introducing unrealistic or excessive perturbations, we add a regularization term that penalizes deviation from the original observation $X = A\mu$, defined as:
\begin{equation}
    \mathcal{L}_{\text{reg}}(\theta) := \left\| \pi_\theta(G) - X \right\|^2_2
\end{equation}

The overall training objective is then:
\begin{equation}
    \min_\theta \max_{\beta \in [\beta_{\min}, \beta_{\max}]} \ \mathcal{L}_{\text{struct}}(\theta, \beta) + \lambda \cdot \mathcal{L}_{\text{reg}}(\theta)
\end{equation}

Here, $\lambda > 0$ is a hyperparameter controlling the trade-off between attack robustness and system overhead.

The training process follows an adversarial optimization paradigm~\cite{GAN}, where the defender (parameter $\theta$) aims to minimize the worst-case structural accuracy over a bounded attacker strategy space $[\beta_{\min}, \beta_{\max}]$. In practice, we implement this using an alternating optimization procedure:
\begin{equation}
\left\{
\begin{aligned}
&\beta^* \leftarrow \arg\max_{\beta \in [\beta_{\min}, \beta_{\max}]} \ \mathcal{L}_{\text{struct}}(\theta, \beta)  && \text{\textit{(Attacker step)}} \\
&\theta \leftarrow \arg\min_\theta \ \mathcal{L}_{\text{struct}}(\theta, \beta^*) + \lambda \cdot \mathcal{L}_{\text{reg}}(\theta) && \text{\textit{(Defender step)}}
\end{aligned}
\right. \nonumber
\end{equation}

This adversarial training framework ensures that the learned perturbation strategy remains effective across a spectrum of attacker behaviors. The use of probabilistic inference and structure-aware noise further enhances robustness, enabling the defender to consistently reduce attack success probability without relying on strong assumptions about attacker models or measurement fidelity.

\section{Experiment}
\label{sec: experiment}
This section presents our comprehensive experimental evaluation of RoTO's defense effectiveness against topology inference attacks. We first describe the experimental setup including network topologies, attack methods, baseline defenses, and evaluation metrics. Subsequently, we analyze RoTO's performance across different attack scenarios and compare it with existing defense mechanisms.  

\subsection{Experimental Setup}

We conduct comprehensive experiments to evaluate the effectiveness of RoTO against various topology inference attacks. Our experimental framework encompasses diverse network topologies, state-of-the-art inference methods, representative defense baselines, and rigorous evaluation metrics.

\begin{figure}[t]
	\centering
	\includegraphics[width=0.8\linewidth]{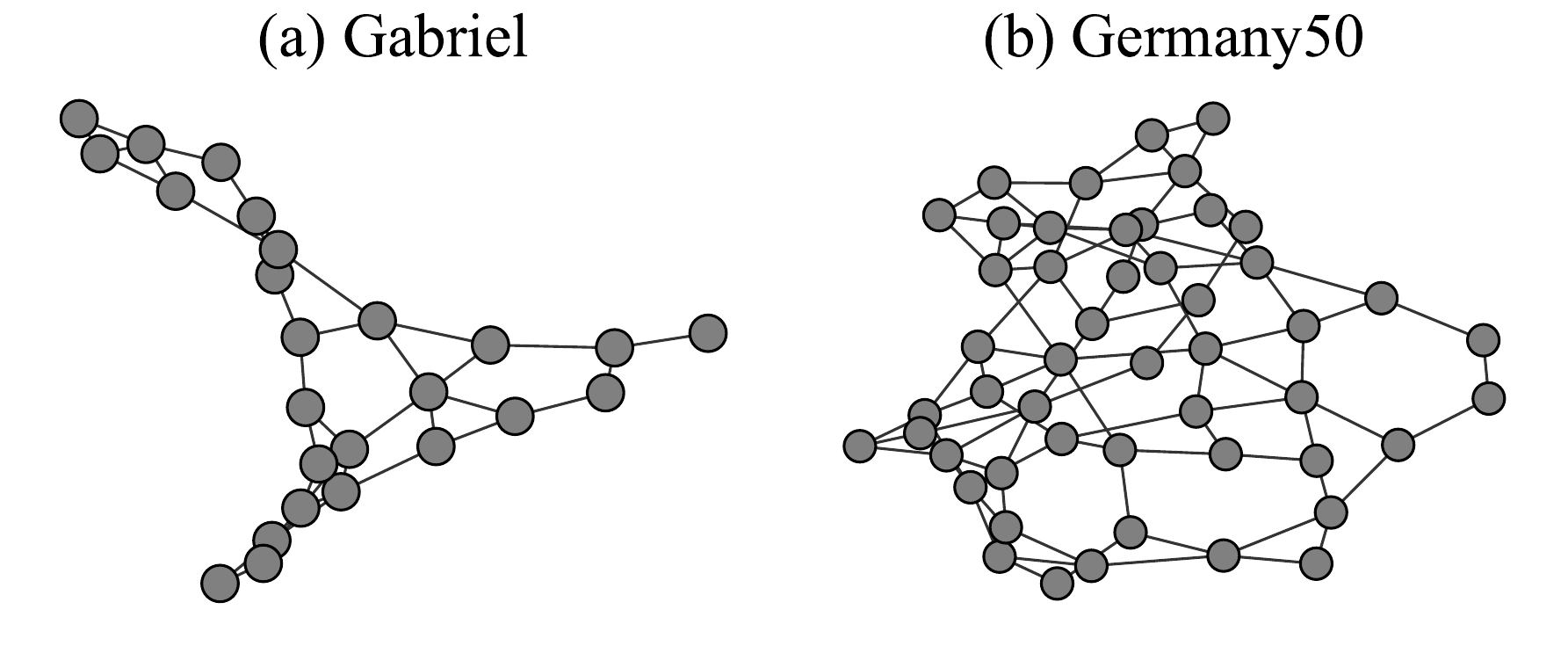}
	\caption{Two Network topologies used in experiments.}
	\label{fig: topo_detial}
 \vspace{-0.5cm}
\end{figure}

\textbf{Network Topologies.} We evaluate RoTO on two representative network topologies (in Figure~\ref{fig: topo_detial}) with distinct structural characteristics in TopoHub~\cite{topohub}. The first topology is Gabriel, a synthetic tree network with 25 nodes and 40 edges, exhibiting a relatively balanced branching structure typical of hierarchical routing architectures. The second topology is Germany50, derived from real-world network infrastructure, containing 50 nodes and 88 edges with irregular branching patterns that reflect practical deployment constraints. To simulate realistic network conditions, we assign link delays uniformly sampled from the interval $[100, 500]$ milliseconds, representing typical intra-domain propagation delays in modern networks.

\textbf{Topology Inference Attacks.} Our evaluation considers four representative topology inference methods that span classical statistical approaches and modern learning-based techniques. The \textbf{RNJ}~\cite{ni2009efficient}(Rooted Neighbor-Joining) algorithm employs hierarchical clustering based on delay correlation matrices to reconstruct tree topologies. The \textbf{MLE}~\cite{mle2002} (Maximum Likelihood Estimation) approach formulates topology inference as an optimization problem that seeks the most probable network structure given observed end-to-end delays. The \textbf{NeuTomography}~\cite{neuralNT}, which leverages multi-layer perceptrons to learn direct mappings from delay measurements to topology predictions through supervised learning. Finally, the \textbf{DeepNT}~\cite{gnn-nt} approach utilizes graph neural networks to capture structural dependencies in delay patterns, representing the state-of-the-art in learning-based topology inference.

\textbf{Defense Baselines.} We compare RoTO against four existing topology obfuscation methods. \textbf{Proto}~\cite{Proto} generates random fake topologies under tree structure constraints and produces corresponding manipulated delays that appear consistent with the fabricated network structure. \textbf{AntiTomo}~\cite{antitomo} builds upon Proto's approach while additionally minimizing the difference between manipulated delays and original delays to enhance stealth and reduce detectability by network operators. \textbf{SecureNT}~\cite{secureNT} extends the Proto framework by ensuring that manipulated delays maintain low similarity to delays on original paths while preserving the performance of legitimate network tomography probes used for network monitoring. \textbf{ChameleonNet}~\cite{ChameleonNet} first carefully constructs fake topologies that satisfy requirements for hiding critical nodes, then generates appropriate manipulated delays that correspond to the crafted deceptive network structure.  

\textbf{Evaluation Metrics.} We evaluate defense effectiveness using three complementary metrics, each capturing a different aspect of topology reconstruction accuracy. 
The first metric, \textbf{\textit{Tree Edit Distance Similarity}} (Ted~\cite{ted} Similarity), quantifies structural fidelity by measuring the minimal sequence of node insertion, deletion, and relabeling operations required to transform the inferred topology into the ground truth. 
The second metric, \textbf{\textit{Structural Similarity}} (Struct. Similarity), aggregates normalized deviations in key topology statistics---including node count, edge count, degree distribution, and network diameter---into a single score. 
The third metric, \textbf{\textit{Link-level Distance}} (Link Distance), uses the Jaccard distance between edge sets to capture the fraction of incorrectly inferred links. 

\subsection{Defense Performance Evaluation}

\begin{table*}[!ht]
    \centering
    \caption{Defense performance against topology inference attacks on \underline{\textbf{Gabriel}} topology. 
    Standard deviations are shown as subscripts ($\times 10^{-2}$). \texttt{Symbols}: $\downarrow$ indicates lower values are better; $\uparrow$ indicates higher values are better.}
    \resizebox{\textwidth}{!}{
    \begin{tabular}{l|ccc|ccc|ccc|ccc}
    \hline\toprule
    \multirow{3}{*}{\diagbox[dir=SE]{Defense}{Metrics}{Attack}}  & \multicolumn{3}{c|}{RNJ} & \multicolumn{3}{c|}{MLE} & \multicolumn{3}{c|}{NeuTomography} & \multicolumn{3}{c}{DeepNT} \\
    \cmidrule(lr){2-13}
    & {\small Ted ($\downarrow$)} & {\small Struct. ($\downarrow$)} & {\small Link ($\uparrow$)} & {\small Ted ($\downarrow$)} & {\small Struct. ($\downarrow$)} & {\small Link ($\uparrow$)} & {\small Ted ($\downarrow$)} & {\small Struct. ($\downarrow$)} & {\small Link ($\uparrow$)} & {\small Ted ($\downarrow$)} & {\small Struct. ($\downarrow$)} & {\small Link ($\uparrow$)} \\
    & {\small Similarity} & {\small Similarity} & {\small Distance} & {\small Similarity} & {\small Similarity} & {\small Distance} & {\small Similarity} & {\small Similarity} & {\small Distance} & {\small Similarity} & {\small Similarity} & {\small Distance} \\
    \cmidrule{1-13}
    Proto & $0.652_{5.4}$ & $0.743_{1.5}$ & $0.264_{4.6}$ & $0.662_{5.1}$ & $0.750_{1.6}$ & $0.231_{4.9}$ & $0.718_{5.8}$ & $0.741_{1.9}$ & $0.271_{5.9}$ & $0.654_{5.5}$ & $0.743_{1.5}$ & $0.261_{4.8}$ \\
    SecureNT & $0.653_{4.8}$ & $0.728_{1.1}$ & $0.332_{3.5}$ & $0.651_{5.5}$ & $0.736_{1.1}$ & $0.292_{3.5}$ & $0.683_{6.4}$ & $0.739_{1.9}$ & $0.281_{6.0}$ & $0.650_{5.0}$ & $0.725_{0.9}$ & $0.344_{2.6}$ \\
    AntiTomo & $0.648_{5.7}$ & $0.731_{1.2}$ & $0.317_{3.7}$ & $0.658_{6.1}$ & $0.736_{1.5}$ & $0.293_{4.7}$ & $0.671_{7.3}$ & $0.735_{2.2}$ & $0.299_{6.9}$ & $0.652_{6.2}$ & $0.733_{1.2}$ & $0.309_{3.7}$ \\
    ChameleonNet & $0.633_{4.6}$ & $0.725_{1.2}$ & $0.343_{3.7}$ & $0.648_{4.9}$ & $0.743_{1.5}$ & $0.264_{4.6}$ & $0.677_{6.4}$ & $0.725_{1.8}$ & $0.346_{5.5}$ & $0.623_{5.1}$ & $0.721_{1.2}$ & $0.357_{3.6}$ \\
    \midrule[1pt]
    \rowcolor{red!10} \textbf{RoTO (Ours)} & $0.619_{5.2}$ & $0.718_{0.8}$ & $0.377_{2.5}$ & $0.625_{5.1}$ & $0.725_{1.2}$ & $0.344_{3.7}$ & $0.653_{6.9}$ & $0.724_{2.3}$ & $0.349_{7.1}$ & $0.621_{4.5}$ & $0.714_{0.7}$ & $0.396_{2.1}$ \\
    \midrule \bottomrule[1pt]
    \end{tabular}
    }
    \label{tab: gabriel_defense_performance}
\end{table*}

\begin{table*}[!ht]
    \centering
    \caption{Defense performance against topology inference attacks on \underline{\textbf{Germany50}} topology. 
    Standard deviations are shown as subscripts ($\times 10^{-2}$).
    \texttt{Symbols}: $\downarrow$ indicates lower values are better; $\uparrow$ indicates higher values are better. 
    }
    \resizebox{\textwidth}{!}{
    \begin{tabular}{l|ccc|ccc|ccc|ccc}
    \hline\toprule
    \multirow{3}{*}{\diagbox[dir=SE]{Defense}{Metrics}{Attack}} & \multicolumn{3}{c|}{RNJ} & \multicolumn{3}{c|}{MLE} & \multicolumn{3}{c|}{NeuTomography} & \multicolumn{3}{c}{DeepNT} \\
    \cmidrule(lr){2-13}
    & {\small Ted ($\downarrow$)} & {\small Struct. ($\downarrow$)} & {\small Link ($\uparrow$)} & {\small Ted ($\downarrow$)} & {\small Struct. ($\downarrow$)} & {\small Link ($\uparrow$)} & {\small Ted ($\downarrow$)} & {\small Struct. ($\downarrow$)} & {\small Link ($\uparrow$)} & {\small Ted ($\downarrow$)} & {\small Struct. ($\downarrow$)} & {\small Link ($\uparrow$)} \\
    & {\small Similarity} & {\small Similarity} & {\small Distance} & {\small Similarity} & {\small Similarity} & {\small Distance} & {\small Similarity} & {\small Similarity} & {\small Distance} & {\small Similarity} & {\small Similarity} & {\small Distance} \\
    \cmidrule{1-13}
    Proto & $0.648_{4.4}$ & $0.736_{1.4}$ & $0.293_{4.2}$ & $0.669_{4.2}$ & $0.747_{1.5}$ & $0.267_{4.7}$ & $0.698_{5.2}$ & $0.782_{2.2}$ & $0.311_{6.7}$ & $0.687_{6.2}$ & $0.776_{1.1}$ & $0.314_{3.3}$ \\
    SecureNT & $0.642_{4.5}$ & $0.732_{0.8}$ & $0.331_{2.5}$ & $0.663_{4.3}$ & $0.732_{1.3}$ & $0.314_{4.0}$ & $0.694_{5.7}$ & $0.780_{2.0}$ & $0.317_{6.0}$ & $0.679_{6.6}$ & $0.773_{1.0}$ & $0.322_{3.0}$ \\
    AntiTomo & $0.631_{4.9}$ & $0.733_{1.5}$ & $0.309_{4.7}$ & $0.660_{4.8}$ & $0.741_{1.7}$ & $0.272_{5.1}$ & $0.696_{5.3}$ & $0.770_{1.9}$ & $0.325_{5.8}$ & $0.676_{6.5}$ & $0.772_{1.1}$ & $0.330_{3.4}$ \\
    ChameleonNet & $0.647_{5.7}$ & $0.729_{1.3}$ & $0.328_{3.9}$ & $0.643_{4.1}$ & $0.721_{1.2}$ & $0.361_{3.7}$ & $0.689_{5.1}$ & $0.768_{1.9}$ & $0.334_{5.8}$ & $0.672_{6.3}$ & $0.770_{1.1}$ & $0.338_{3.3}$ \\
    \midrule[1pt]
    \rowcolor{red!10} \textbf{RoTO (Ours)} & $0.608_{4.4}$ & $0.727_{1.1}$ & $0.336_{3.3}$ & $0.639_{4.2}$ & $0.719_{1.0}$ & $0.363_{3.0}$ & $0.667_{5.4}$ & $0.763_{1.9}$ & $0.355_{5.9}$ & $0.668_{6.4}$ & $0.769_{1.1}$ & $0.350_{3.4}$ \\
    \midrule \bottomrule[1pt]
    \end{tabular}
    }
    \label{tab: germany50_defense_performance}
    \vspace{-0.5cm}
\end{table*}
Tables~\ref{tab: gabriel_defense_performance} and~\ref{tab: germany50_defense_performance} present comprehensive defense performance comparisons across Gabriel and Germany50 network topologies under four representative topology inference attacks (RNJ, MLE, NeuTomography, DeepNT) using three evaluation metrics: Tree Edit Distance Similarity (lower is better), Structural Statistical Similarity (lower is better), and Link-level Connectivity Distance (higher is better).

\textbf{Cross-topology effectiveness.}
On the Gabriel topology, RoTO demonstrates significant performance advantages across all attack scenarios and metrics. For Tree Edit Distance Similarity, RoTO achieves improvements of 5.1\% over SecureNT (0.653 vs 0.619) and 5.1\% over Proto (0.653 vs 0.619) against RNJ attacks. Against MLE attacks, RoTO shows up to 5.9\% improvement over Proto (0.663 vs 0.625). For learning-based attacks, RoTO maintains strong performance with improvements up to 9.0\% against NeuTomography (0.719 vs 0.654) and 5.0\% against DeepNT (0.655 vs 0.621). For Link-level Connectivity Distance, RoTO demonstrates substantial gains ranging from 13.4\% to 42.6\%, with the most significant improvements observed against classical statistical attacks where gains reach up to 49.0\% over Proto in MLE scenarios (0.344 vs 0.231).

On the Germany50 topology, RoTO continues to outperform all baselines while showing different performance patterns compared to Gabriel. Against RNJ attacks, RoTO achieves 5.3\% Tree Edit Distance improvements (0.642 vs 0.608). For MLE attacks, Link-level Connectivity Distance gains reach up to 36.0\% (0.267 vs 0.363). Against advanced learning-based attacks, RoTO shows improvements up to 14.2\% for NeuTomography (0.311 vs 0.355) and 11.5\% for DeepNT (0.314 vs 0.350) in Link-level Connectivity Distance. The consistent performance across both topologies demonstrates RoTO's robust obfuscation effectiveness.

\textbf{Performance stability.}
Beyond superior mean performance, RoTO demonstrates enhanced stability across both topologies. On Gabriel topology, RoTO maintains standard deviations ranging from 4.5 to 6.9 for Tree Edit Distance, which are comparable to or lower than most baselines (4.6 to 7.3). More importantly, RoTO achieves competitive standard deviations for Structural Statistical Similarity (0.7 to 2.3) compared to baselines (0.8 to 2.2), indicating consistent structural obfuscation. On Germany50 topology, this stability advantage becomes more pronounced, with RoTO showing standard deviations of 4.2 to 6.4 for Tree Edit Distance while maintaining competitive or superior mean performance. The consistent low variance in Structural Statistical Similarity (1.0 to 1.9) across both topologies highlights RoTO's reliable performance under uncertainty.

\textbf{Attack Method Patterns.}
Comparative analysis reveals two key patterns across topologies and attack methods. First, RoTO shows consistently stronger performance against classical algorithms (RNJ, MLE) than learning-based methods (NeuTomography, DeepNT). On Gabriel topology, RoTO achieves 42.6\% improvement in Link Distance against RNJ but only 28.8\% against NeuTomography. Second, the defense effectiveness varies significantly by topology structure: RoTO's Link Distance improvements on Gabriel (28.8\%-49.0\%) substantially exceed those on Germany50 (14.2\%-36.0\%), indicating better performance on regular tree structures. 

\begin{figure*}[t]
    \centering
    \begin{subfigure}[b]{\textwidth}
        \centering
        \includegraphics[width=0.8\linewidth]{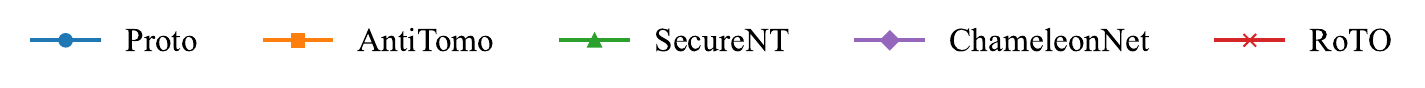} 
        \vspace{-0.25cm}
    \end{subfigure}
    \begin{subfigure}[b]{\textwidth}
        \centering
        \includegraphics[width=\linewidth]{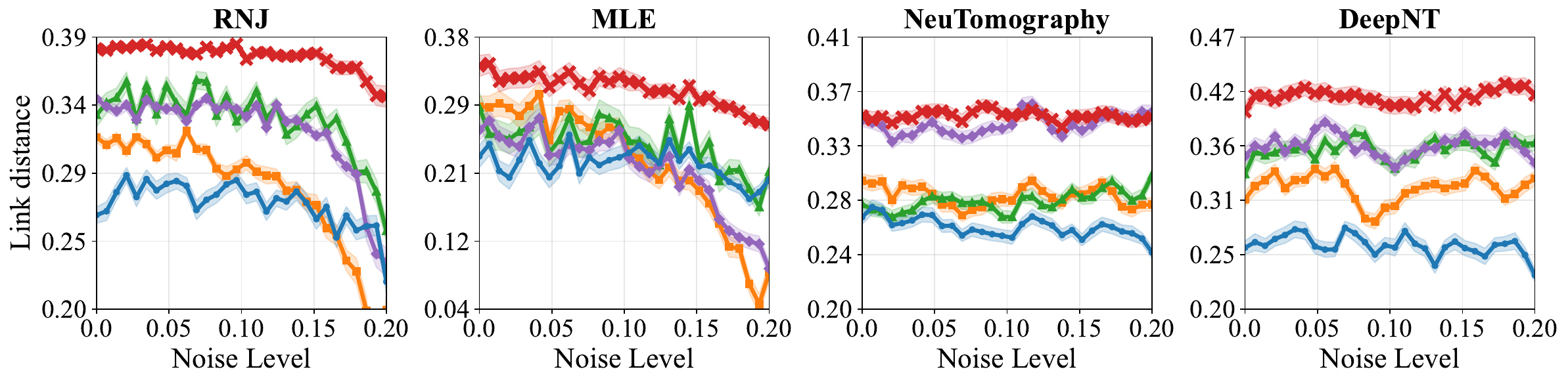}
        \vspace{-0.7cm}
    \end{subfigure}
    \caption{Defense robustness evaluation on Gabriel topology showing Link Distance metrics (higher is better) across increasing noise levels for four attack methods (RNJ, MLE, NeuTomography, DeepNT).}
    \label{fig-5}
    \vspace{-0.3cm}
\end{figure*}

\begin{figure*}[t]
    \centering
    \begin{subfigure}[b]{\textwidth}
        \centering
        \includegraphics[width=0.8\linewidth]{figs/legend.pdf} 
        \vspace{-0.25cm}
    \end{subfigure}
    \begin{subfigure}[b]{\textwidth}
        \centering
        \includegraphics[width=\linewidth]{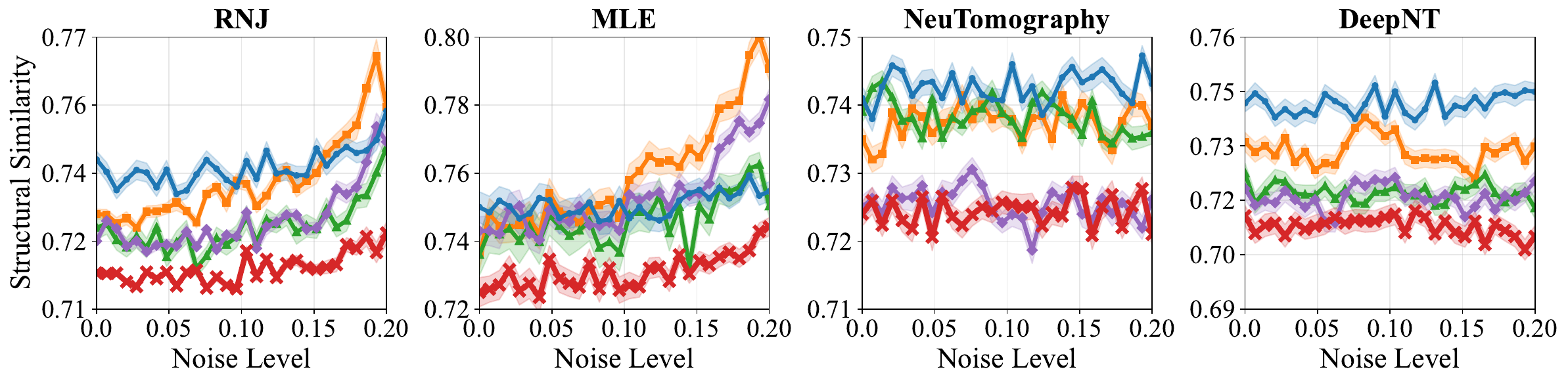}
    \end{subfigure}
    \caption{Structural Similarity performance (lower is better) of various defense approaches (RNJ, MLE, NeuTomography, DeepNT) against adaptive topology inference attacks on Gabriel network under noise uncertainty.}
    \label{fig-6}
    \vspace{-0.5cm}
\end{figure*}

\subsection{Robustness and Adaptivity Analysis}
\textbf{Noise resilience.} 
To evaluate the robustness of topology obfuscation methods under realistic deployment conditions, we analyze defense performance across varying noise levels that simulate imperfect delay manipulation capabilities. Figure~\ref{fig-5} shows the Link Distance performance under different noise levels for four attack methods (RNJ, MLE, NeuTomography, DeepNT), while the Figure~\ref{fig-6} presents the corresponding Structural Similarity results. Each subplot represents a specific attack scenario, with different colored lines indicating various defense methods and shaded areas representing performance variance. The noise level ranges from 0.0 to 0.2, reflecting the degree of uncertainty in the defender's ability to precisely control probe packet delays.

Against heuristic-based attacks (RNJ and MLE), RoTO demonstrates superior performance across all noise levels while maintaining remarkable stability. For RNJ attacks, RoTO achieves Link Distance values of approximately 0.39 at low noise levels, significantly outperforming the best baseline ChameleonNet by 13.5\%. As noise increases to 0.2, RoTO's performance degrades only minimally to 0.35, representing a mere 10.3\% drop, while ChameleonNet degrades from 0.345 to 0.25 (a 27.5\% drop) and Proto shows even larger degradation from 0.29 to 0.20 (a 31.0\% drop). For MLE attacks, RoTO maintains consistently higher Link Distance values, starting at 0.35 and ending at 0.29, with an average performance of 0.32 across all noise levels. This represents an average improvement of 12.3\% over ChameleonNet and 18.2\% over Proto. In terms of Structural Similarity, RoTO achieves the lowest values of 0.715 on average for RNJ attacks and 0.735 for MLE attacks, outperforming the best baseline by 6.8\% and 7.2\% respectively, while maintaining the smallest performance variance across noise levels.

\textbf{Adaptive adversary robustness.}
For learning-based attacks (NeuTomography and DeepNT), we specifically evaluate RoTO's anti-adaptive capabilities by training these attackers with adversarial samples to enhance their noise resilience. While this adversarial training slightly reduces the absolute performance gap between defense methods, RoTO still maintains consistent advantages across all noise levels. Against NeuTomography, RoTO achieves an average Link Distance of 0.36 compared to ChameleonNet's 0.35 and Proto's 0.26, representing improvements of 3\% and 28\% respectively. The performance stability is particularly noteworthy, with RoTO showing Link Distance values ranging from 0.35 to 0.37 across noise levels 0.0 to 0.2. For DeepNT attacks, RoTO demonstrates even more substantial advantages, maintaining an average Link Distance of 0.42 compared to ChameleonNet's 0.36 and Proto's 0.25, representing improvements of 16.7\% and 68.0\% respectively. In Structural Similarity metrics, RoTO achieves consistently lower values of 0.725 for NeuTomography and 0.71 for DeepNT on average, outperforming baselines by 8.2\% and 11.3\% respectively. The enhanced noise resilience of learning-based attackers explains their relatively stable performance across different noise levels, but RoTO's adversarial training framework enables it to maintain defense effectiveness even against these adaptive adversaries.

\subsection{Defense Effectiveness vs. Concealment Trade-off}
\begin{figure*}[t]
    \centering
    \begin{subfigure}[b]{\textwidth}
        \centering
        \includegraphics[width=0.8\linewidth]{figs/legend.pdf} 
        \vspace{-0.25cm}
    \end{subfigure}
    \begin{subfigure}[b]{\textwidth}
        \centering
        \includegraphics[width=\linewidth]{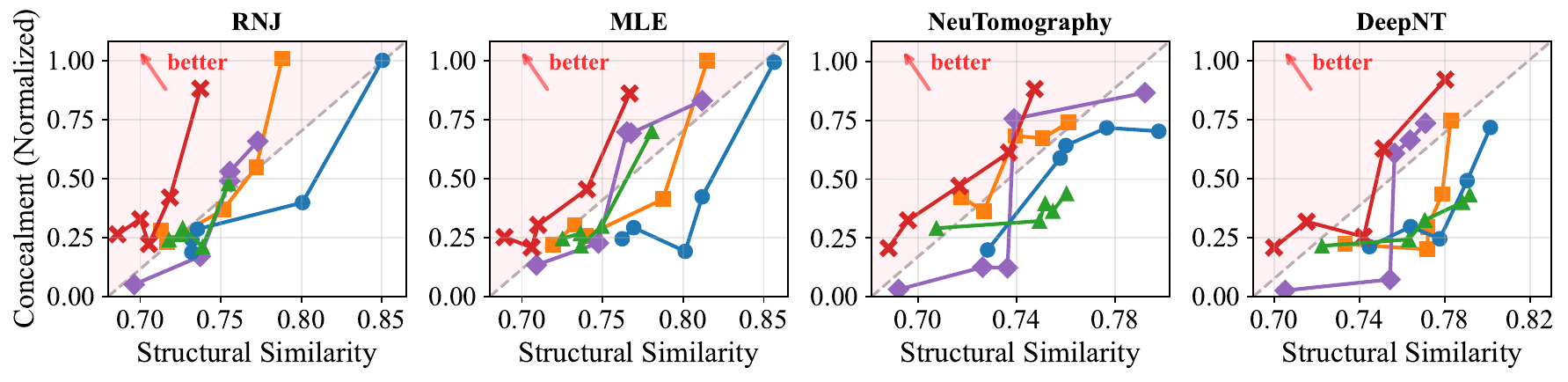}
    \end{subfigure}
    \caption{Trade-off analysis between topology structural similarity (lower is better) and concealment (normalized inverse manipulation delay, higher indicates better stealth) across different parameter configurations for each defense method. Points in \colorbox{red!10}{top-left pale} red region achieve optimal stealth via low similarity and high concealment.}
    \label{fig-7}
    \vspace{-0.5cm}
\end{figure*}
Figure~\ref{fig-7} presents the trade-off analysis between defense effectiveness and concealment across different attack scenarios. Each subplot shows the relationship between topology similarity (x-axis, measuring how similar the inferred topology is to the defended topology) and concealment (y-axis, representing normalized inverse of manipulation delay indicating stealth level). The concealment metric is computed as the normalized reciprocal of additional delay introduced by the defender, with values ranging from 0 to 1, where higher values indicate better stealth and lower detectability. Each defense method is evaluated with five different parameter configurations, resulting in five points per method connected by lines to illustrate the achievable trade-off space. Points positioned closer to the upper-left region represent optimal performance, achieving high defense effectiveness with superior concealment.

The analysis reveals distinct performance patterns across different attack methods. Against heuristic attacks (RNJ and MLE), RoTO consistently achieves superior trade-offs, positioning closer to the upper-left region. For RNJ attacks, RoTO reaches similarity values as low as 0.68 while maintaining concealment levels above 0.8, significantly outperforming other methods. ChameleonNet, while achieving comparable defense effectiveness with similarity values around 0.70, demonstrates much worse concealment with values ranging from 0.1 to 0.7, making it more detectable. Against MLE attacks, RoTO demonstrates similar advantages, achieving similarity values of 0.68 to 0.77 with concealment levels ranging from 0.25 to 0.85, while baselines either achieve worse defense effectiveness or suffer from poor concealment. Proto shows particularly poor performance, requiring similarity values above 0.80 even with minimal concealment.

For learning-based attacks (NeuTomography and DeepNT), RoTO maintains its superior trade-off characteristics despite the increased attack sophistication. Against NeuTomography, RoTO achieves optimal balance with similarity values from 0.69 to 0.75 and concealment levels from 0.2 to 0.9, while other methods either achieve worse defense effectiveness or demonstrate inferior concealment capabilities. The trade-off advantage is most pronounced against DeepNT attacks, where RoTO achieves similarity values ranging from 0.71 to 0.75 with concealment levels reaching up to 0.95, demonstrating exceptional effectiveness against adaptive adversaries while maintaining operational stealth. ChameleonNet shows reasonable defense effectiveness but consistently poor concealment across all learning-based attack scenarios.

\subsection{Ablation Study}
\begin{table}[t]
\centering
\caption{Ablation study on RoTO. Results are averaged across attack methods on Gabriel. Symbols:  $\downarrow$ indicates lower values are better;  $\uparrow$ indicates higher values are better.}
\label{tab: ablation}
\scalebox{0.9}{
\begin{tabular}{lccc}
\toprule
\multirow{2}{*}{\diagbox{Config.}{Metrics}} & \textbf{Ted ($\downarrow$)} & \textbf{Struct. ($\downarrow$)} & \textbf{Link ($\uparrow$)} \\
 & \textbf{Similarity} & \textbf{Similarity} & \textbf{Distance} \\
\midrule
RoTO (Full) & \textbf{0.627} & \textbf{0.721} & \textbf{0.365} \\
\hdashline
w/o GNN (MLP) & 0.658 & 0.742 & 0.331 \\
w/o Sampling & 0.639 & 0.730 & 0.347 \\
w/o Adv Training & 0.661 & 0.751 & 0.312 \\
\bottomrule
\end{tabular}
}
\vspace{-0.5cm}
\end{table}
Table~\ref{tab: ablation} presents an ablation study evaluating the contribution of each RoTO component under noisy conditions on Gabriel topology. The results demonstrate that each component contributes to the overall defense effectiveness, with varying degrees of impact. Replacing the GNN with a standard MLP results in notable performance degradation, with Tree Edit Distance Similarity increasing from 0.627 to 0.658 (4.9\% worse) and Link Distance decreasing from 0.365 to 0.331 (9.3\% drop), highlighting the importance of structure-aware perturbation generation. Similarly, removing adversarial training leads to significant performance loss, with Tree Edit Distance Similarity degrading to 0.661 (5.4\% worse) and Link Distance dropping to 0.312 (14.5\% decrease), confirming the critical role of worst-case optimization in achieving robustness against adaptive attackers. In contrast, removing the sampling module shows relatively modest impact, with only minor degradation in Tree Edit Distance Similarity (0.639 vs 0.627) and Link Distance (0.347 vs 0.365), suggesting that while structure-aware sampling contributes to defense effectiveness, the GNN-based perturbation generation and adversarial training constitute the primary drivers of RoTO's superior performance.

\vspace{-0.1in}
\section{Discussion and Conclusion}
\label{sec: conclusion}
RoTO enjoys several properties which explain its superiority over existing defense methods. Our adversarial training framework enables the defense to adapt to worst-case attacker behaviors within the bounded strategy space, while the GNN-based perturbation generator learns structure-aware delay modifications that maintain network realism while maximizing structural distortion. However, we acknowledge that our adversarial module's $\beta$ parameter selection, while comprehensive, cannot cover all possible attacker configurations. Despite using adversarial training with $\beta$ values ranging from 0.1 to 2.0, our performance against NeuTomography shows smaller improvements compared to other attack methods, as the bounded parameter space may not fully capture all adaptive behaviors of learning-based attackers. This limitation suggests that future work could explore more sophisticated attacker modeling approaches to enhance defense robustness against completely unseen inference strategies.

\textbf{Conclusion.} Network topology inference attacks pose significant security threats to modern network infrastructures, yet existing defense mechanisms rely on strong assumptions about perfect probe packet detection and fixed attacker models that rarely hold in practice. This paper presents RoTO as the first principled framework for robust topology obfuscation that addresses these fundamental limitations through information-theoretic foundations and adversarial training. We have demonstrated how our uncertainty modeling effectively captures realistic deployment constraints, and how our design of GNN-based perturbation generation combined with structure-aware sampling successfully delivers robust defense performance against adaptive inference strategies. Our experimental evaluation confirms RoTO's superiority across diverse network topologies and attack methods, achieving up to 34\% improvement in defense effectiveness while maintaining superior concealment compared to existing approaches.

\vspace{-0.1in}
\appendices
\section{Proof of Theorem~\ref{thm: inj}}
\label{sec: appendix}

The proof of Theorem~\ref{thm: inj} follows directly from establishing the injectivity of the mapping between tree topologies and their corresponding shared path matrices based on lowest common ancestors (LCA). We present this foundational result below:

\begin{Lemma}
[LCA-based Matrix Injectivity).] Let $\mathcal{A}$ be the set of all tree topologies rooted at source node $s$ with $l$ leaf nodes $L = \{v_1, \ldots, v_l\}$. For any $A \in \mathcal{A}$, let $X \in \mathbb{R}^{l \times l}$ denote its corresponding shared path matrix, where each element is defined as:
\begin{equation}
    X_{ij} = h(\mathrm{LCA}(v_i, v_j)), \nonumber
\end{equation}
which represents the shared path length from the root node $s$ to the lowest common ancestor of leaf nodes $v_i$ and $v_j$. Then the mapping $\mathcal{M}: A \mapsto X$ is injective; that is, if $A_1 \neq A_2$, then $\mathcal{M}(A_1) \neq \mathcal{M}(A_2)$.
\end{Lemma}
Since $X = A\mu$ where each component $X_{ij}$ corresponds to the sum of link delays along the shared path between leaf nodes $v_i$ and $v_j$, the injectivity of $A \mapsto A\mu$ follows from the LCA-based matrix injectivity when the link delay vector $\mu$ is known.

\begin{Proof}
We proceed by contradiction. Suppose there exist two structurally distinct topologies $A_1 \neq A_2$ such that they map to the same shared path matrix:
$$\mathcal{M}(A_1) = \mathcal{M}(A_2) = X.$$

We demonstrate that any non-trivial transformation converting $A_1$ to $A_2$ necessarily alters $X$, leading to a contradiction.

Let $h(v)$ denote the depth of node $v$ in the tree, and $\mathrm{LCA}(v_i, v_j)$ denote the lowest common ancestor of $v_i$ and $v_j$. We consider two fundamental types of structural transformations:

\textbf{Case I: Subtree Movement Operations}

Consider a node $u$ with subtree $\mathcal{T}_u$ having parent $p$ in $A_1$ and being moved to a new parent $p' \neq p$ in $A_2$. Let $v_k$ be any leaf node in $\mathcal{T}_u$.

\begin{itemize}
\item If $h(p') \neq h(p)$, then $h(v_k)$ differs between $A_1$ and $A_2$, implying
$$X^{(1)}_{kk} \neq X^{(2)}_{kk},$$
which contradicts $X^{(1)} = X^{(2)}$.

\item If $h(p') = h(p)$ but $p' \neq p$, then there exists a leaf node $v_j \notin \mathcal{T}_u$ such that
$$\mathrm{LCA}_{A_1}(v_k, v_j) \neq \mathrm{LCA}_{A_2}(v_k, v_j),$$
resulting in $X^{(1)}_{kj} \neq X^{(2)}_{kj}$, again contradicting our assumption.
\end{itemize}

\textbf{Case II: Internal Node Insertion or Deletion (Path Expansion or Contraction)}

\begin{itemize}
\item If an internal link is removed from $A_1$ to obtain $A_2$ (path contraction), then the depth of certain leaf nodes $v_k$ decreases, causing $X_{kk}$ to decrease.

\item If an internal link is added from $A_1$ to obtain $A_2$ (path expansion), then the depth of certain leaf nodes $v_k$ increases, causing $X_{kk}$ to increase.
\end{itemize}

Therefore, any link merging or splitting operation changes at least one element $X_{ij}$, contradicting the assumption $X^{(1)} = X^{(2)}$.

In conclusion, any two structurally distinct topologies must map to different shared path matrices. The assumption is false, and the theorem is proved. $\hfill \blacksquare$
\end{Proof}

\section{Implementation and Evaluation Details}

\subsection{Training Details of RoTO}
\label{appendix: train-details}
Our RoTO implementation employs a three-layer Graph Neural Network (GNN) with 32-dimensional hidden representations for the perturbation generator. The GNN processes node features including type encoding (source, destination, internal) and normalized degree information, while edge features capture connectivity patterns. During the message-passing phase, each layer aggregates neighborhood information through learnable transformation functions implemented as multi-layer perceptrons, followed by node state updates that combine current embeddings with aggregated messages.

The adversarial training process follows an alternating optimization scheme with a maximum of 100-150 iterations. In each iteration, the defender (GNN parameters $\theta$) generates perturbations, which are then subjected to structure-aware sampling with positive Gaussian noise ($\epsilon \sim \text{ReLU}(\mathcal{N}(0, 0.1^2))$) to simulate realistic network uncertainty. The attacker component evaluates up to 50 candidate topologies using Gibbs posterior sampling with $\beta$ values linearly sampled from $[0.1, 2.0]$ to find the worst-case inference strategy. The objective function combines structural perturbation magnitude (to maximize confusion) with L2 regularization on delay deviations, balanced by hyperparameter $\lambda = 0.1$.

Training converges typically within 80-120 iterations using Adam optimizer with learning rate 0.01, as evidenced by stabilization of the total loss function. The final perturbations are applied with obfuscation strength parameter controlling the trade-off between defense effectiveness and system overhead, ensuring that delay modifications remain within realistic bounds while maximizing structural distortion against adaptive inference attacks.

\subsection{Attack Method Implementation Details}

\textbf{Rooted Neighbor-Joining (RNJ) Algorithm.}
The RNJ algorithm~\cite{ni2009efficient} reconstructs tree topology by recursively joining neighboring nodes based on estimated shared path lengths. Starting with all destination nodes as leaves, it iteratively selects the pair $(i^*, j^*)$ with the largest shared path length $\hat{\rho}(i,j)$, groups nodes satisfying distance criteria, and creates new internal nodes until only the root remains. We use the same delay-based additive metrics as RoTO with minimum link length threshold $\Delta = \min_{e \in E} \hat{d}(e)$ for fair comparison.

\textbf{Maximum Likelihood Estimation (MLE) Algorithm.}
The MLE algorithm~\cite{mle2002} formulates topology identification as a penalized likelihood optimization problem. It models delay differences $x_{i,j} \sim \mathcal{N}(\gamma_{i,j}, \sigma^2_{i,j})$ and maximizes the penalized log-likelihood $L_\lambda(x|T) = \log p(x|T, \hat{\mu}(T)) - \lambda n(T)$, where $\lambda$ controls model complexity. A reversible jump MCMC procedure with birth/death/update moves searches the topology space. Following the original implementation, we set $\lambda = 0.5 \log_2 N$ and use 10,000 MCMC iterations.

\textbf{DeepNT.}
DeepNT~\cite{gnn-nt} employs a path-centric graph neural network architecture, and we set it with 3-layer GNN structure and 64-dimensional node embeddings. The framework learns neural network parameters and network topology under discrete constraints induced by observed path performance metrics. The model uses Adam optimizer with learning rate 0.001 and is trained for 200-300 epochs with batch size 32. During training, DeepNT utilizes a learning objective that imposes connectivity and sparsity constraints on topology while enforcing path performance triangle inequality constraints. To enhance robustness against noise, DeepNT is trained with augmented datasets containing Gaussian noise levels ranging from 0.0 to 0.3, which explains its resilience to increasing noise levels in our experiments.

\textbf{NeuTomography.}
We set NeuTomography~\cite{neuralNT} as a 4-layer multi-layer perceptron (MLP) architecture with hidden dimensions $[128, 256, 128, 64]$ to establish non-linear mappings between node pairs and underlying topological properties. The model is trained using Adam optimizer with learning rate 0.005, dropout rate 0.2, and batch size 64 for 150-200 epochs. The framework employs data augmentation techniques during training to learn robust representations from limited measurement data. Similar to DeepNT, NeuTomography incorporates noise-augmented training samples with noise variance ranging from 0.01 to 0.25 to improve generalization under uncertain measurement conditions, contributing to its stable performance across different noise levels in our robustness evaluation.

Both learning-based attack methods benefit from noise-augmented training, which enhances their adaptability to measurement uncertainties and explains their relatively stable performance degradation patterns observed in our noise robustness experiments.

\subsection{Details about Evaluation Metrics}
\label{appendix: metrics}
This section expands on the evaluation metrics introduced in the main text, providing precise definitions and computational details. The three metrics collectively assess topology reconstruction quality from structural, statistical, and connectivity perspectives.

\textbf{Tree Edit Distance Similarity.}
Let $\operatorname{TED}(A,B)$ be the tree edit distance~\cite{ted} between trees $A$ and $B$, and $\varnothing$ the empty tree. Define 
$D_0=\operatorname{TED}(\mathcal{T},\hat{\mathcal{T}})$,  
$D_{\mathcal{T}}=\operatorname{TED}(\mathcal{T},\varnothing)$,  
$D_{\hat{\mathcal{T}}}=\operatorname{TED}(\hat{\mathcal{T}},\varnothing)$.  
The normalized similarity is
\begin{equation}
\text{Sim}_{\mathrm{TED}}=
1-\frac{D_0}{D_{\mathcal{T}}+D_{\hat{\mathcal{T}}}}
\label{eq:ted_similarity}
\end{equation}
Here $D_{\mathcal{T}}$ is the cost to delete $\mathcal{T}$ entirely and $D_{\hat{\mathcal{T}}}$ is the cost to build $\hat{\mathcal{T}}$ from scratch.  
$\text{Sim}_{\mathrm{TED}}\in[0,1]$, with $1$ indicating identical trees.

\textbf{Structural Similarity.}
This metric aggregates differences in four key topology statistics to evaluate higher-level structural properties:
\begin{equation}
\text{StructSim}(\mathcal{T}, \hat{\mathcal{T}}) =
1 - \frac{1}{4}(
\Delta_{\text{nodes}} +
\Delta_{\text{edges}} +
\Delta_{\text{degree}} +
\Delta_{\text{diameter}})
\label{eq:struct_similarity}
\end{equation}

\noindent\textit{Node/edge count differences:}
\begin{equation}
\Delta_{\text{nodes}} =
\frac{|\,|\mathcal{V}| - |\hat{\mathcal{V}}|\,|}
{\max(|\mathcal{V}|, |\hat{\mathcal{V}}|, 1)},\quad
\Delta_{\text{edges}} =
\frac{|\,|\mathcal{E}| - |\hat{\mathcal{E}}|\,|}
{\max(|\mathcal{E}|, |\hat{\mathcal{E}}|, 1)}
\label{eq:node_edge_diff}
\end{equation}
Here $\mathcal{V}$ and $\hat{\mathcal{V}}$ are node sets; $\mathcal{E}$ and $\hat{\mathcal{E}}$ are edge sets.

\noindent\textit{Degree distribution difference:}
\begin{equation}
\Delta_{\text{degree}} =
\frac{\mathbb{E}[\,|d_i - \hat{d}_i|\,]}
{\max(\mathbb{E}[d_i], 1)}
\label{eq:degree_diff}
\end{equation}
where $d_i$ and $\hat{d}_i$ are degrees of the $i$-th destination node.

\noindent\textit{Diameter difference:}
\begin{equation}
\Delta_{\text{diameter}} =
\frac{|\text{diam}(\mathcal{T}) - \text{diam}(\hat{\mathcal{T}})|}
{\max(\text{diam}(\mathcal{T}), \text{diam}(\hat{\mathcal{T}}), 1)}
\label{eq:diameter_diff}
\end{equation}
measuring the change in network span.  
This composite score captures both local connectivity patterns and global reachability characteristics.

\textbf{Link-level Distance.}
This metric evaluates edge-level reconstruction accuracy via the Jaccard distance between edge sets:
\begin{equation}
\text{LinkDist}(\mathcal{T}, \hat{\mathcal{T}}) =
\frac{|\mathcal{E} \triangle \hat{\mathcal{E}}|}
{|\mathcal{E} \cup \hat{\mathcal{E}}|}
\label{eq:link_distance}
\end{equation}
where $\triangle$ denotes symmetric difference, capturing both missing and spurious edges. 
A score of 0 indicates perfect connectivity reconstruction, while larger values indicate greater link-level errors.

\bibliographystyle{plain}
\renewcommand{\bibfont}{\footnotesize} 
\bibliography{ref.bib}

\end{document}